\documentclass{article}
\usepackage{spconf,amsmath,graphicx,hyperref}
\usepackage{makecell}
\usepackage{multirow}
\usepackage{booktabs}
\usepackage{amsfonts}
\usepackage{cite}
\usepackage{bm}

\usepackage[subtle,title=normal,sections=normal,margins=normal,mathdisplays=normal,mathspacing=normal, bibliography=normal]{savetrees}
\title{I-DCCRN-VAE: An Improved Deep Representation Learning Framework for Complex VAE-based Single-channel Speech Enhancement}
\name{Jiatong Li, Simon Doclo\thanks{This work was funded by the Deutsche Forschungsgemeinschaft (DFG, German Research Foundation) under Germany’s Excellence Strategy - EXC 2177/1 - Project ID 390895286 and Project ID 352015383 - SFB 1330 B2.}}
\address{Dept. of Medical Physics and Acoustics and Cluster of Excellence Hearing4all, \\ Carl von Ossietzky Universität Oldenburg, Germany \\ jiatong.li@uni-oldenburg.de, simon.doclo@uni-oldenburg.de}
\begin{document}
\ninept
\maketitle
\begin{abstract}
    Recently, a complex variational autoencoder (VAE)-based single-channel speech enhancement system based on the DCCRN architecture has been proposed. In this system, a noise suppression VAE (NSVAE) learns to extract clean speech representations from noisy speech using pretrained clean speech and noise VAEs with skip connections. In this paper, we improve DCCRN-VAE by incorporating three key modifications: 1) removing the skip connections in the pretrained VAEs to encourage more informative speech and noise latent representations; 2) using $\beta$-VAE in pretraining to better balance reconstruction and latent space regularization; and 3) a NSVAE generating both speech and noise latent representations. Experiments show that the proposed system achieves comparable performance as the DCCRN and DCCRN-VAE baselines on the matched DNS3 dataset but outperforms the baselines on mismatched datasets (WSJ0-QUT, Voicebank-DEMEND), demonstrating improved generalization ability. In addition, an ablation study shows that a similar performance can be achieved with classical fine-tuning instead of adversarial training, resulting in a simpler training pipeline.
\end{abstract}
\begin{keywords}
Variational Autoencoder, Single-channel speech enhancement, Latent representations
\end{keywords}
\section{Introduction}
\label{sec:intro}
Recently, several generative models, e.g. based on the variational autoencoders (VAEs) \cite{fang2021variational,bie2022unsupervised,xiang2022bayesian,xiang2022deep,xiang2023two,sadeghi2023fast,xiang2024deep,li2025investi}, generative adversarial networks \cite{pascual17_interspeech,baby2019sergan,phan2020improving,wali2022generative,10508391}, and diffusion models \cite{lu2022conditional,richter2023speech,nortier2024unsupervised,gonzalez2024investigating}, have been proposed for speech enhancement.  
VAEs consist of an encoder-decoder architecture, where the encoder maps the input data into latent representations, conditioned by a latent regularization loss, 
and the decoder aims at reconstructing data from these representations \cite{Kingma2014}. Since the VAE framework facilitates efficient posterior inference and reliable reconstruction, several VAE-based approaches have been proposed for single-channel speech enhancement. For example, the Bayesian permutation training (PVAE) system  \cite{xiang2022bayesian, xiang2022deep} uses a noise suppression VAE (NSVAE) to learn the latent representations of two pretrained VAEs, one for clean speech (CVAE) and one for noise (NVAE). Several improvements were proposed for this system. In \cite{li2025investi}, we showed that removing the latent regularization loss for the pretrained VAEs improves performance and generalization. In addition, in \cite{xiang2023two} it was proposed to use adversarial training to fine-tune the CVAE and NVAE decoders. Since the PVAE only estimates the clean speech magnitude in the Short-time Fourier Transform (STFT) domain in combination with the noisy phase, in \cite{xiang2024deep}, the PVAE was extended to the complex domain based on the DCCRN architecture, leading to DCCRN-VAE. Contrary to the PVAE, the DCCRN-VAE employs skip connections for the pretrained VAEs. Its NSVAE encoder only generates speech latent representations, and only the CVAE decoder is fine-tuned with the NSVAE encoder and skip connections using adversarial training. 

\begin{figure}[t!]
\begin{minipage}[b]{0.59\linewidth}
  \centering
  \centerline{\includegraphics[width=4.55cm]{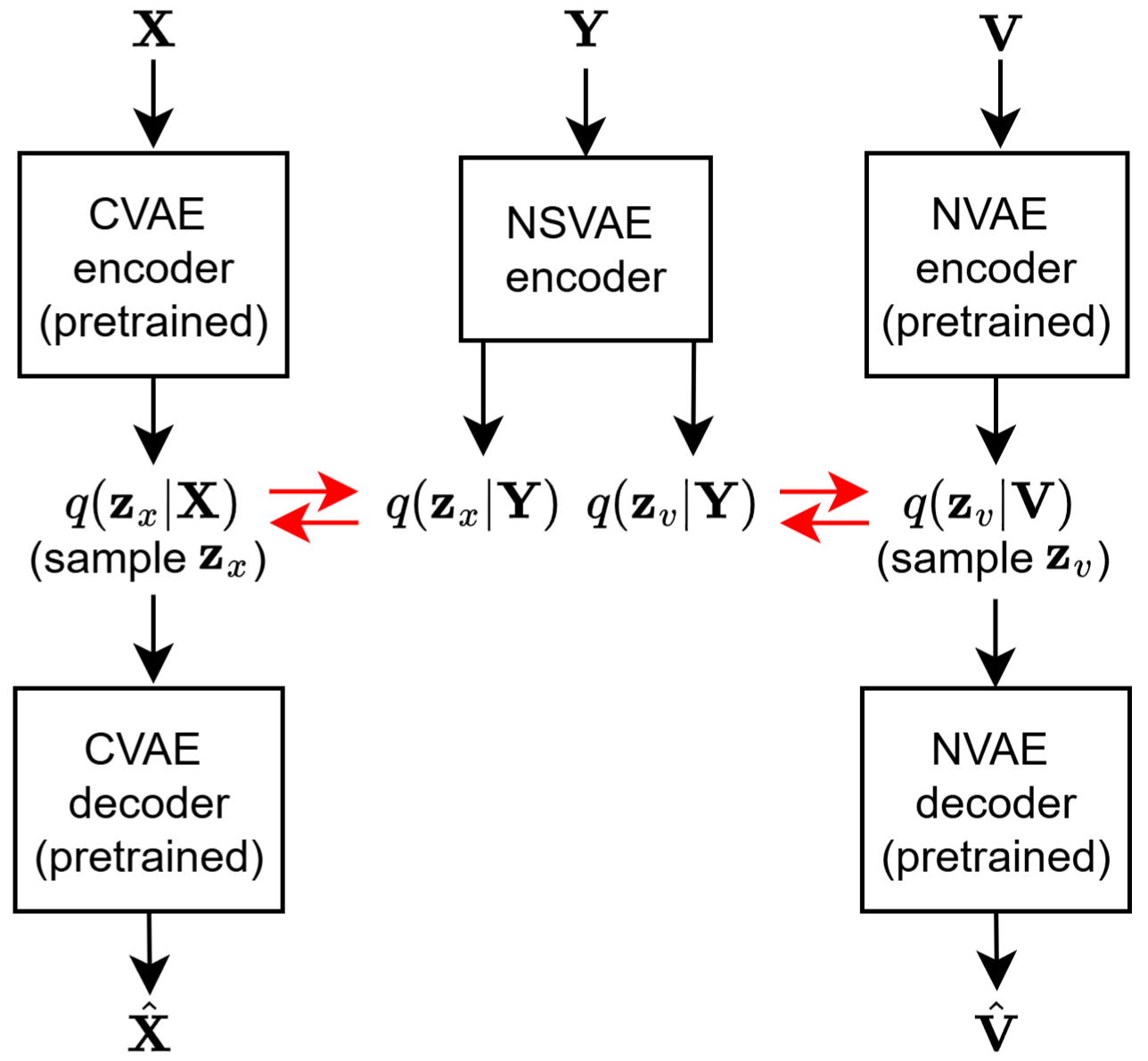}}
  \centerline{(a) NSVAE encoder training}
  \label{figa}
\end{minipage}
\hfill
\begin{minipage}[b]{0.40\linewidth}
  \centering
  \centerline{\includegraphics[width=2.48cm]{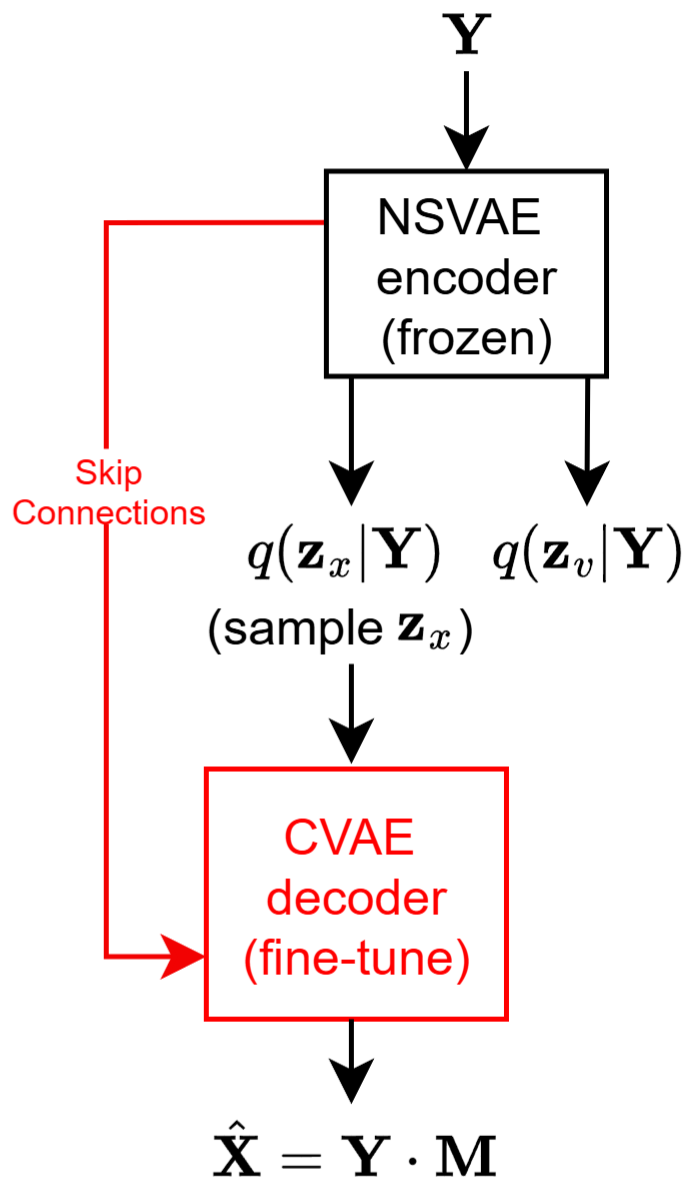}}
  \centerline{(b) Decoder Fine-tuning}
    \label{figb}
\end{minipage}
 \caption{Overview of the improved DCCRN-VAE, consisting of two pretrained VAEs without skip connections, i.e. clean speech VAE (CVAE) and noise VAE (NVAE), and the noise suppression VAE (NSVAE).} 
\label{fig:diagram}
\end{figure}

In this paper, we propose an improved complex VAE-based model, the I-DCCRN-VAE (see Fig. \ref{fig:diagram}), by introducing three key modifications to the DCCRN-VAE: 1) we remove the skip connections in the pretrained VAEs, as they can dominate the reconstruction process and make the latent representations less informative; 2) inspired by our previous work \cite{li2025investi}, we use $\beta$-VAE in pretraining to better balance reconstruction and latent space regularization; and
3) Similarly to the PVAE system, the NSVAE encoder generates both speech and noise latent representations.
In our experiments, we train the proposed I-DCCRN-VAE and the baseline DCCRN-VAE on the DNS3 challenge dataset. Fine-tuning the CVAE decoder for both systems either uses classical fine-tuning or adversarial training. The results show that the proposed I-DCCRN-VAE achieves comparable speech enhancement performance as the baseline DCCRN-VAE (and DCCRN) on the matched dataset, but outperforms the baselines on two mismatched datasets (WSJ0-QUT, Voicebank-DEMAND).
Notably, the I-DCCRN-VAE achieves this improved generalization without adversarial training, which is a significant advantage, as it simplifies training by avoiding the convergence and sensitivity issues common to adversarial methods.
\section{The Proposed I-DCCRN-VAE system}
\label{sec:format}
After introducing the signal model, in this section, we describe the proposed I-DCCRN-VAE system, highlighting the differences with the baseline DCCRN-VAE \cite{xiang2024deep} in terms of pretrained VAEs, noise suppression VAE, and decoder fine-tuning.
\subsection{Signal Model}
In the STFT domain, the observed noisy speech vector $\mathbf{Y}_n \in \mathbb{C}^F$ at time frame $n \in [1, N]$, where $N$ and $F$ denote the number of time frames and frequency bins, is given by
\begin{equation*} 
    \mathbf{Y}_n = \mathbf{X}_n + \mathbf{V}_n, \tag{1}
\end{equation*}
where $\mathbf{X}_n \in \mathbb{C}^F$ and $\mathbf{V}_n \in \mathbb{C}^F$ denote the clean speech and noise vectors. In the following, the time frame index $n$ is omitted for simplicity, except when it is required explicitly.

We assume $\mathbf{X}$ and $\mathbf{V}$ to be generated from random processes involving latent speech and noise representations $\mathbf{z}_x \in \mathbb{C}^L$ and $\mathbf{z}_v \in \mathbb{C}^L$, $L$ denotes the latent dimension. These random processes are described by the likelihoods $p_{\scriptstyle\theta_x}(\mathbf{X}|\mathbf{z}_x)$ and $p_{\scriptstyle \theta_v}(\mathbf{V}|\mathbf{z}_v)$. $\mathbf{z}_x$ and $\mathbf{z}_v$ can be sampled from the posterior distributions $q_{\scriptstyle \phi_x}(\mathbf{z}_x|\mathbf{X})$ and $q_{\scriptstyle \phi_v}(\mathbf{z}_v|\mathbf{V})$. Assuming that the above-mentioned distributions are estimated by VAEs, $\phi_x$ and $\theta_x$ denote the encoder and decoder parameters of the clean speech VAE (CVAE), while $\phi_v$ and $\theta_v$ denote the encoder and decoder parameters of the noise VAE (NVAE). The prior distributions for the latent representations $\mathbf{z}_x$ and $\mathbf{z}_v$ are denoted as $p(\mathbf{z}_x)$ and $p(\mathbf{z}_v)$.
Assuming $\mathbf{z}_x$ and $\mathbf{z}_v$ to be independent, $\mathbf{z}_x$ and $\mathbf{z}_v$ can also be sampled from the noisy posterior distribution $q_{\scriptstyle \phi_y}(\mathbf{z}_x, \mathbf{z}_v|\mathbf{Y})\hspace{-0.05cm}=\hspace{-0.05cm}q_{\scriptstyle \phi_y}(\mathbf{z}_x|\mathbf{Y})q_{\scriptstyle \phi_y}(\mathbf{z}_v|\mathbf{Y})$, where $\phi_y$ denotes the encoder parameters of a noise suppression VAE (NSVAE). In the following, the encoder and decoder parameters are omitted for simplicity. 
\subsection{System Description}
Similar to the DCCRN-VAE system \cite{xiang2024deep}, the proposed I-DCCRN-VAE system in Fig. \ref{fig:diagram} consists of two pretrained VAEs, a clean speech VAE (CVAE) and a noise VAE (NVAE), and a noise suppression VAE (NSVAE). The training process consists of three steps: 1) pretraining the CVAE and NVAE using clean speech $\mathbf{X}$ and noise $\mathbf{V}$, 2) training the NSVAE encoder to extract speech and noise representations $\mathbf{z}_{x}$ and $\mathbf{z}_{v}$ from noisy speech $\mathbf{Y}$, and 3) fine-tuning the CVAE decoder for better speech enhancement. The I-DCCRN-VAE differs from the DCCRN-VAE in the pretraining and NSVAE training steps.

\textbf{1) Pretrained VAEs}: 
Unlike the DCCRN-VAE, which uses skip connections for the CVAE and the NVAE, we don't consider skip connections for the pretrained VAEs in the I-DCCRN-VAE. This forces all information to pass through the latent bottleneck, encouraging to learn more informative speech and noise latent representations rather than relying on skip connections for reconstruction. In addition, we also use $\beta$-VAE \cite{higgins2017betavae} to control the balance between reconstruction and latent space regularization. The pretraining loss for the CVAE is given by:
\begin{align*}
& -\mathbb{E}_{q(\mathbf{z}_x|\mathbf{X})}[\log p(\mathbf{X}\vert \mathbf{z}_{x})]\ \hspace{-0.1cm}+ \beta \textrm{KL}(q(\mathbf{z}_{x}\vert \mathbf{X})\Vert p(\mathbf{z}_{x})),\tag{2} \label{cvaeloss}
\end{align*}
where $\mathbb{E}$ denotes expectation, $\textrm{KL}(\cdot\Vert\cdot)$ denotes the Kullback–Leibler (KL) divergence and $\beta$ denotes the KL weight factor. As in \cite{xiang2024deep}, the posterior distribution is assumed to be a complex multivariate Gaussian distribution with a diagonal covariance matrix and relation matrix, i.e.
\begin{align*}
   & {q({\bf z}_x|{\bf{X}})} = \mathcal{N} \left({{\bm{\mu}}_{x},\operatorname{diag}(\bm{\sigma}_{x}), \operatorname{diag}(\bm{\delta}_{x})}\right), \tag{3}\label{cvae_post_distri}
\end{align*}
where the mean, variance and relation vectors, ${\bm\mu}_{x} \in \mathbb{C}^L$, ${\bm\sigma}_{{x}} \in \mathbb{R}_+^L$ and $\bm{\delta}_{{x}} \in \mathbb{C}^L$, are the outputs of the CVAE encoder. The prior distribution $p(\mathbf{z}_{x})$ is assumed to be a complex multivariate standard Gaussian distribution, $p(\mathbf{z}_{x})=\mathcal{N} ({{\bf{0}},{\bf{I}}, {\bf{0}}})$, where $\bf{I}$ denotes the identity matrix. To allow for backpropagation, the reparameterization trick \cite{nakashika2020complex} is used to sample $\mathbf{z}_x$ from $q(\mathbf{z}_x|\mathbf{X})$. To improve reconstruction, the first term in (\ref{cvaeloss}) is replaced by a combined loss on the complex and magnitude spectrograms, i.e.
\begin{align*}
   &\frac{1}{N}\sum_{n=1}^N \left(\Vert\mathbf{X}_n-\hat{\mathbf{X}}_n\Vert_2^2+\Vert\vert\mathbf{X}_n\vert-\vert\hat{\mathbf{X}}_n\vert\Vert_2^2\right), \tag{4}\label{cvae_recon}
\end{align*}
where $\hat{\mathbf{X}}_n$ denotes the estimated clean speech STFT vector and $\vert \cdot \vert$ denotes the magnitude of a vector (element-wise).
The NVAE assumes a similar loss and similar distributions as the CVAE, which is not explained in detail here.

\textbf{2) Noise suppression VAE (Fig. \ref{fig:diagram}(a))}: In contrast to the NSVAE encoder in the DCCRN-VAE, which generates only speech representations $\mathbf{z}_{x}$ from noisy speech $\mathbf{Y}$ and applies a residual loss to align intermediate features between the NSVAE encoder and the pretrained CVAE encoder, the NSVAE encoder in the I-DCCRN-VAE generates both speech and noise representations $\mathbf{z}_{x}$ and $\mathbf{z}_{v}$ without using a residual loss in training. This follows the probabilistic generative modeling derived in \cite{xiang2022bayesian}, which provides a more complete generative basis. Aiming at making the posterior distributions $q(\mathbf{z}_x|\mathbf{Y})$ and $q(\mathbf{z}_v|\mathbf{Y})$ from the NSVAE encoder similar to the posterior distributions $q(\mathbf{z}_x|\mathbf{X})$ and $q(\mathbf{z}_v|\mathbf{V})$ from the pretrained VAEs, the NSVAE is trained by minimizing the loss   
 \begin{align*}
  \textrm{KL}\left({q({\bf z}_x|{\bf{Y}})}||{q({\bf z}_x|{\bf{X}})}\right) + \alpha\textrm{KL}\left({q({\bf z}_v|{\bf{Y}})}||{q({\bf z}_v|{\bf{V}})}\right),
  \tag{5}\label{nsvaeloss}
\end{align*}
where $\alpha$ denotes the noise latent weight factor. It should be noted that when $\alpha=0$, the NSVAE is trained to generate only speech representations $\mathbf{z}_{x}$.
Similar to (\ref{cvae_post_distri}), the posterior distributions estimated from the NSVAE encoder are assumed to follow a complex multivariate Gaussian distribution, i.e.
\begin{align*}
   & {q({\bf z}_x|{\bf{Y}})} = \mathcal{N} \left({{\bm\mu}_{{yx}},\operatorname{diag}({\bm\sigma}_{{yx}}), \operatorname{diag}({\bm\delta}_{{yx}})}\right), \tag{6}\\
   & {q({\bf z}_v|{\bf{Y}})} = \mathcal{N}\left({{\bm\mu}_{{yv}},\operatorname{diag}({\bm\sigma}_{{yv}}), \operatorname{diag}({\bm\delta}_{{yv}})}\right),\tag{7}
\label{nsvaeposterior}
\end{align*} 
where the mean vectors ${\bm\mu}_{{yx}}$ and ${\bm\mu}_{{yv}}$, the variance vectors ${\bm\sigma}_{{yx}}$ and ${\bm\sigma}_{{yv}}$, and the relation vectors ${\bm\delta}_{{yx}}$ and ${\bm\delta}_{{yv}}$ are the outputs of the NSVAE encoder.

\textbf{3) CVAE decoder fine-tuning (Fig. \ref{fig:diagram}(b))}: As estimation errors in the posterior distribution $q({\bf z}_x|{\bf{Y}})$ degrade the speech enhancement performance, it was proposed in \cite{xiang2024deep} to fine-tune the CVAE decoder while keeping the NSVAE encoder frozen. Skip connections were added in fine-tuning to provide the CVAE decoder with detailed encoder features and to combat the vanishing gradient problem. Similarly as in \cite{xiang2024deep}, the CVAE decoder in the I-DCCRN-VAE is fine-tuned to generate the complex mask $\mathbf{M}$, which is used to estimate the clean speech $\hat{\mathbf{X}}$ as
\begin{align*}
   & \hat{\mathbf{X}}=\mathbf{Y}\cdot \mathbf{M}, \tag{8}\label{estspeech}
\end{align*}
where the multiplication is performed element-wise. Fine-tuning is performed using the Scale Invariant Signal-to-Distortion Ratio (SI-SDR) loss between the estimated speech $\hat{\mathbf{x}}$ and the clean speech $\mathbf{x}$ in the time domain obtained by inverse STFT and overlap-add, i.e.
\begin{equation}
    \mathcal{L}_{\text{SI-SDR}} = -10 \log_{10} \left( \frac{\left\| \mathbf{x}_d \right\|_2^2}{\left\| \mathbf{x}_d  - \hat{\mathbf{x}} \right\|_2^2} \right),    \mathbf{x}_d=\frac{\langle \hat{\mathbf{x}}, \mathbf{x} \rangle}{\|\mathbf{x}\|_2^2}\mathbf{x}. \tag{9}\label{sisdrloss}
\end{equation}
For the fine-tuning step, various training schemes can be applied. Besides classical fine-tuning, which only minimizes SI-SDR loss in (\ref{sisdrloss}), adversarial training has been used in \cite{xiang2024deep}, involving a discriminator network. The discriminator learns to distinguish between estimated clean speech and true clean speech, thereby encouraging the model to produce more realistic results.

\section{Experiments}
\label{sec:experiments}
This section first presents the experimental setup, including the training and evaluation datasets, the network structure, and the training procedure. Then, the experimental results are presented and discussed, evaluating key differences between the proposed I-DCCRN-VAE and the baseline DCCRN-VAE. 
\subsection{Training and Evaluation Datasets}
\label{ssec:datasets}
To train all considered VAE-based speech enhancement systems, we used anechoic clean speech and noise from the DNS3 dataset \cite{reddy21_interspeech}, sampled at 16kHz. It should be noted that for clean speech, we only considered the read speech (leaving out emotional speech), while for noise, we did not consider the DEMAND dataset, since it was used for evaluation. We randomly split 50$\%$ of speakers for CVAE pretraining, 40\% of speakers for NSVAE training and fine-tuning, and 10\% of speakers for validation. The noise data was split similarly. For NSVAE training and fine-tuning, noisy speech was generated by the DNS script at signal-to-noise ratios (SNRs) between -10\,dB and 15\,dB. In total, we generated 30 hours of data for pretraining, 20 hours of data for NSVAE training and CVAE decoder fine-tuning, and 10 hours of data for validation. 

To evaluate the speech enhancement performance, we used three datasets. As the matched evaluation dataset, we used the official synthetic DNS3 test set at SNRs between 0\,dB and 19\,dB. To test the generalization ability, we used two mismatched datasets with different speakers and noise from the training dataset, namely WSJ0-QUT \cite{bie2022unsupervised} and VoiceBank-DEMAND (VB-DMD) \cite{valentinibotinhao16_interspeech}. WSJ0-QUT includes cafe, home, street and car noise at SNRs of -5\,dB, 0\,dB and 5\,dB. The official VB-DMD test set includes room, office, bus, cafe and public square noise at SNRs of 2.5\,dB, 7.5\,dB, 12.5\,dB and 17.5\,dB.
\subsection{Network and Training}
\label{ssec:network}
We used a similar STFT framework and network architectures as for the DCCRN-VAE system \cite{xiang2024deep}.
The time-domain signals are transformed to the STFT domain using a Hann window with a frame length of 400, 25\% overlap, and a FFT length of 512. 
For all VAEs, the dimension of the latent representations is equal to $L=128$. The CVAE and NVAE encoders contain six Conv2d blocks and one complex LSTM layer. The channels for the Conv2d blocks are [32, 64, 128, 128, 256, 256], with a kernel size of (5,2) and a stride of (2,1). The complex LSTM layer outputs the $L$-dimensional mean, variance and relation vectors: (${\bm\mu}_{{x}}$, ${\bm\sigma}_{{x}}$ and ${\bm\delta}_{{x}}$ for the CVAE; ${\bm\mu}_{{v}}$, ${\bm\sigma}_{{v}}$ and ${\bm\delta}_{{v}}$ for the NVAE). The NSVAE encoder has a similar structure, where the only difference is that the LSTM layer generates both speech and noise vectors (${\bm\mu}_{{yx}}$, ${\bm\mu}_{{yv}}$, ${\bm\sigma}_{{yx}}$, ${\bm\sigma}_{{yv}}$, ${\bm\delta}_{{yx}}$, ${\bm\delta}_{{yv}}$). The CVAE and NVAE decoders mirror their respective encoders in reverse order.
When adversarial training is used for the CVAE decoder fine-tuning, the discriminator has a similar structure to the CVAE encoder, including six Conv2d blocks and one real LSTM layer with a single output.

All networks were trained for a maximum of 1000 epochs. The training was stopped early if the validation loss did not decrease for 20 consecutive epochs. The Adam optimizer was used with a learning rate of 3e-4 (CVAE, NVAE, NSVAE) and a learning rate of 8e-5 (discriminator for adversarial training). All learning rates were halved if the validation loss did not improve for 3 consecutive epochs. The batch size was set to 15. The code can be found on \url{https://github.com/iris1997jiatong/I-DCCRN-VAE}.
\subsection{Experimental Results: Hyperparameter Optimization}
\label{ssec:results}
Aiming at finding the optimal configuration set of hyperparameters for the proposed I-DCCRN-VAE, in the first set of experiments, we evaluate key differences with the DCCRN-VAE. More in particular, we investigate the influence of skip connections and latent space regularization in the pretrained VAEs as well as the influence of the NSVAE training target. It should be noted that in this set of experiments, we only consider classical fine-tuning for the CVAE decoder fine-tuning.
\begin{table}[t]
\vspace{-6pt}
\caption{{Average reconstruction SI-SDR (dB) and KL loss (KLL) between posterior distributions and prior distributions of pretrained CVAE and NVAE with or without skip connections (SC) using different KL weight factor $\beta$ for DNS3 dataset.}}
\label{tab: sc_pre}
\vspace{1mm}
\setlength{\tabcolsep}{7pt} 
\centering
\normalsize 
\begin{tabular}{c|c|cc|cc}
	\toprule
	\multirow{2}{*}{} & \multirow{2}*{$\beta$} & \multicolumn{2}{c|}{CVAE} & \multicolumn{2}{c}{NVAE} \\
	& & \multicolumn{1}{c}{SI-SDR} & \multicolumn{1}{c|}{KLL} & \multicolumn{1}{c}{SI-SDR} & \multicolumn{1}{c}{KLL} \\
    \cline{2-6}
	\hline
    \multirow{4}{*}{\makecell{\scriptsize{Without}\\\scriptsize{SC}}}
    & $0.001$  & 15.7 & 303.9 & 16.6 & 398.1 \\
    &$0.01$  & 14.7 & 67.3 & 14.9 & 114.7 \\

    & $0.1$  & 13.0 & 24.0 & 12.4 & 41.6 \\
    & $1$  & 8.4 & 7.7 & 5.5 & 11.6 \\
    \hline 
    \multirow{1}{*}{\scriptsize{With SC}} & - & 39.0 & 0.0 &38.1 & 0.0 \\
	\bottomrule                           
\end{tabular}	
\caption{{Average SI-SDR (dB) and PESQ on different datasets, with and without skip connections (SC) and using different KL weight factors $\beta$ in the pretrained VAEs (using the NSVAE trained with $\alpha=1$).}}
\label{tab: beta_se}
\setlength{\tabcolsep}{2.7pt} 
\vspace{1mm}
\centering
\begin{tabular}{c|c|cc|cc|cc}
	\toprule
	 & \multirow{2}*{$\beta$} & \multicolumn{2}{c|}{DNS3} & \multicolumn{2}{c|}{WSJ0-QUT} &  \multicolumn{2}{c}{VB-DMD} \\
	& & \multicolumn{1}{c}{SI-SDR} & \multicolumn{1}{c|}{PESQ} & \multicolumn{1}{c}{SI-SDR} & \multicolumn{1}{c|}{PESQ} & \multicolumn{1}{c}{SI-SDR} & \multicolumn{1}{c}{PESQ} \\
    \cline{2-7}
	\hline
     \multirow{4}*{\makecell{\scriptsize{Without}\\\scriptsize{SC}}} &$0.001$  & 16.9 & \textbf{2.52} & 8.6 & 1.61 & 17.8 & 2.33 \\
     & $0.01$  & \textbf{17.2} & 2.49 & \textbf{8.7} & \textbf{1.65} & \textbf{18.0} & \textbf{2.44} \\

    &$0.1$  & 16.8 & 2.35 & 8.4 & 1.62 & \textbf{18.0} & 2.43 \\
    &$1$  & 16.0 & 2.23 & 7.4 & 1.55 & 17.2 & 2.32 \\
    \hline
    \scriptsize{With SC} & -  & 11.7 & 1.71 & 0.0 & 1.19 & 14.1 & 2.16 \\
	\bottomrule                           
\end{tabular}	
\caption{{Average SI-SDR (dB) and PESQ on different datasets for different NSVAE training targets (using pretrained VAEs without skip connections with $\beta=0.01$).}}
\label{tab: nsvae_res}
\setlength{\tabcolsep}{2.7pt} 
\centering
\vspace{1mm}
\begin{tabular}{c|cc|cc|cc}
	\toprule
	  \multirow{2}*{$\alpha$} & \multicolumn{2}{c|}{DNS3} & \multicolumn{2}{c|}{WSJ0-QUT} &  \multicolumn{2}{c}{VB-DMD} \\
	    & \multicolumn{1}{c}{SI-SDR} & \multicolumn{1}{c|}{PESQ} & \multicolumn{1}{c}{SI-SDR} & \multicolumn{1}{c|}{PESQ} & \multicolumn{1}{c}{SI-SDR} & \multicolumn{1}{c}{PESQ} \\
    \cline{2-7}
	\hline
    0  & 16.9 & 2.44 & 7.9 & 1.62 & 17.9 & 2.43 \\
     1  & \textbf{17.2} & \textbf{2.49} & \textbf{8.7} & \textbf{1.65} & \textbf{18.0} & \textbf{2.44} \\
	\bottomrule                           
\end{tabular}
\end{table}
\begin{table*}[t!]	
\setlength{\tabcolsep}{5pt}
\centering
\caption{{Average SI-SDR (dB), PESQ and ESTOI (with 95\% confidence interval) of DCCRN, DCCRN-VAE and I-DCCRN-VAE with two different CVAE decoder fine-tuning methods, i.e. classical fine-tuning (CF), and adversarial training (ADV), evaluated on different datasets.}}
\label{tab: ablation_adv}
\centering
\small
\vspace{1mm}
\begin{tabular}{c|ccc|ccc|ccc}
	\toprule
	 \multirow{2}*{System} & \multicolumn{3}{c|}{DNS3} & \multicolumn{3}{c|}{WSJ0-QUT} & \multicolumn{3}{c}{VB-DMD} \\
	 & SI-SDR & PESQ & ESTOI & SI-SDR & PESQ & ESTOI & SI-SDR & PESQ & ESTOI \\
    \cline{1-10}
    Unprocessed  & \makecell{9.1 \\ ($\pm\,$0.9)} & \makecell{1.58 \\ ($\pm\,$0.07)} & \makecell{0.81 \\ ($\pm\,$0.02)} & \makecell{-2.6 \\ ($\pm\,$0.3)} & \makecell{1.14 \\ ($\pm\,$0.01)} & \makecell{0.50 \\ ($\pm\,$0.01)} & \makecell{8.4 \\ ($\pm\,$0.4)} & \makecell{1.97 \\ ($\pm\,$0.05)} & \makecell{0.79 \\ ($\pm\,$0.01)} \\
	\hline 
    (1) DCCRN \cite{hu20g_interspeech}  & \makecell{16.6 \\ ($\pm\,$0.8)} & \textbf{\makecell{2.54 \\ ($\pm\,$0.10)}} & \textbf{\makecell{0.90 \\ ($\pm\,$0.01)}} & \makecell{7.1 \\ ($\pm\,$0.3)} & \makecell{1.59 \\ ($\pm\,$0.03)} & \makecell{0.67 \\ ($\pm\,$0.01)} & \makecell{17.5 \\ ($\pm\,$0.3)} & \makecell{2.38 \\ ($\pm\,$0.04)} & \makecell{0.81 \\ ($\pm\,$0.01)}   \\
	\hline
    (2) DCCRN-VAE (CF) & \makecell{16.8 \\ ($\pm\,$0.7)} & \makecell{2.38 \\ ($\pm\,$0.09)} & \makecell{0.88 \\ ($\pm\,$0.02)} & \makecell{6.8 \\ ($\pm\,$0.3)} & \makecell{1.49 \\ ($\pm\,$0.03)} & \makecell{0.65 \\ ($\pm\,$0.01)} & \makecell{17.1 \\ ($\pm\,$0.3)} & \makecell{2.36 \\ ($\pm\,$0.04)} & \makecell{0.81 \\ ($\pm\,$0.01)} \\
    (3) DCCRN-VAE (ADV) \cite{xiang2024deep}  & \textbf{\makecell{17.8 \\ ($\pm\,$0.7)}} & \makecell{2.50 \\ ($\pm\,$0.09)} & \textbf{\makecell{0.90 \\ ($\pm\,$0.01)}} & \makecell{7.2 \\ ($\pm\,$0.3)} & \makecell{1.54 \\ ($\pm\,$0.03)} & \makecell{0.67 \\ ($\pm\,$0.01)} & \makecell{17.5 \\ ($\pm\,$0.3)} & \makecell{2.37 \\ ($\pm\,$0.04)} & \makecell{0.81 \\ ($\pm\,$0.01)} \\
    \cline{1-10}

    (4) I-DCCRN-VAE (CF) (\textbf{Proposed})  & \makecell{17.2 \\ ($\pm\,$0.7)} & \makecell{2.49 \\ ($\pm\,$0.09)} & \textbf{\makecell{0.90 \\ ($\pm\,$0.01)}} & \makecell{8.7 \\ ($\pm\,$0.3)} & \textbf{\makecell{1.65 \\ ($\pm\,$0.03)}} & \textbf{\makecell{0.70 \\ ($\pm\,$0.01)}} & \makecell{18.0 \\ ($\pm\,$0.3)} & \textbf{\makecell{2.44 \\ ($\pm\,$0.04)}} & \textbf{\makecell{0.83 \\ ($\pm\,$0.01)}} \\

    (5) I-DCCRN-VAE (ADV) (\textbf{Proposed})  & \makecell{17.5 \\ ($\pm\,$0.7)} & \makecell{2.49 \\ ($\pm\,$0.09)} & \textbf{\makecell{0.90 \\ ($\pm\,$0.01)}} & \textbf{\makecell{8.9 \\ ($\pm\,$0.3)}} & \textbf{\makecell{1.65 \\ ($\pm\,$0.03)}} & \textbf{\makecell{0.70 \\ ($\pm\,$0.01)}} & \textbf{\makecell{18.1 \\ ($\pm\,$0.3)}} & \textbf{\makecell{2.44 \\ ($\pm\,$0.04)}} & \textbf{\makecell{0.83 \\ ($\pm\,$0.01)}}  \\

\bottomrule                             
\end{tabular}	
\end{table*}

Table \ref{tab: sc_pre} shows the influence of skip connections and the KL weight factor $\beta$ in (\ref{cvaeloss}) on the reconstruction quality and the latent space of the pretrained CVAE and NVAE for the DNS3 dataset. We use reconstruction SI-SDR to measure reconstruction quality and KL loss (KLL) between estimated posterior distributions ($q(\mathbf{z}_x|\mathbf{X})$, $q(\mathbf{z}_v|\mathbf{V})$) and prior distributions ($p(\mathbf{z}_x)$, $p(\mathbf{z}_v)$) to assess the regularization of the latent space. Lower KL loss indicates a more regularized space. For both CVAE and NVAE, it can be observed that including skip connections yields a much higher reconstruction SI-SDR but a much lower KL loss close to zero than without skip connections. This indicates posterior collapse for the pretrained VAEs, suggesting that speech and noise latent representations are not very informative for speech and noise reconstruction when using skip connections. Without skip connections, it can be observed that as $\beta$ decreases from 1 to 0.001, the reconstruction SI-SDR and the KLL of both CVAE and NVAE increase. This indicates a clear trade-off: decreasing $\beta$ improves reconstruction quality at the cost of a less regularized latent space. 

Table \ref{tab: beta_se} evaluates the influence of skip connections and $\beta$ in the pretrained VAEs on the overall speech enhancement performance in terms of SI-SDR and wide-band Perceptual Evaluation of Speech Quality (PESQ) \cite{941023} for the matched DNS3 dataset and the mismatched datasets. First, it can be observed that including skip connections yields significantly lower SI-SDR and PESQ scores than without skip connections. This can be explained by the less informative latent representations in the pretrained VAEs, which the NSVAE encoder is trained to match. Therefore, the NSVAE encoder fails to learn useful information for speech reconstruction from the pretrained VAEs. Without skip connections, a clear trend can be observed, where SI-SDR and PESQ across all datasets first increase and then decrease as $\beta$ decreases, with $\beta=0.01$ yielding the best performance (except for PESQ on the matched DNS3 dataset). Combined with results in Table \ref{tab: sc_pre}, we may conclude that both pretrained reconstruction quality and latent space regularization affect the speech enhancement performance. As $\beta$ decreases from 1 to 0.01, the improved pretrained reconstruction quality leads to better speech enhancement performance for all considered datasets. However, while $\beta=0.001$ further improves the pretrained reconstruction quality, the speech enhancement performance degrades, especially for both mismatched datasets. This performance drop is likely due to the highly unregularized latent space for $\beta=0.001$, which degrades the generalization ability.

Table \ref{tab: nsvae_res} shows the influence of the NSVAE training target in (\ref{nsvaeloss}) on the speech enhancement performance for different datasets. For $\alpha=0$, the NSVAE encoder only generates speech latent representations, while for $\alpha=1$, the NSVAE encoder generates both speech and noise latent representations. It should be noted that, based on the results in Table \ref{tab: beta_se}, we consider pretrained VAEs without skip connections with $\beta=0.01$ here. The results show that training the NSVAE to generate both speech and noise representations ($\alpha=1$) yields consistently higher SI-SDR and PESQ scores across all datasets compared to generating only speech representations ($\alpha=0$). This suggests that explicitly modeling the noise component contributes to an extraction of speech information from the noisy mixture, thereby improving the speech enhancement performance.

\subsection{Experimental Results: Comparison with Baselines}
In this section, we compare the performance of the proposed I-DCCRN-VAE, using the optimal configuration (without skip connections in pretrained VAEs, $\beta=0.01$, $\alpha=1$), with DCCRN\cite{hu20g_interspeech} and DCCRN-VAE with the residual loss \cite{xiang2024deep}. For the DCCRN-VAE and I-DCCRN-VAE systems, we also compare adversarial training and classical fine-tuning for the CVAE decoder. For a fair comparison, the DCCRN baseline was trained on the same noisy dataset used for NSVAE training and fine-tuning, while the DCCRN-VAE baseline used the same datasets as the I-DCCRN-VAE.

Table \ref{tab: ablation_adv} shows the average SI-SDR, wideband PESQ, and Extended Short-Time Objective Intelligibility (ESTOI) \cite{5495701} scores for the matched DNS3 dataset and the mismatched datasets. First, it can be observed that for all datasets that compared to classical fine-tuning, adversarial training provides a much larger performance benefit for the baseline DCCRN-VAE (systems (2) and (3)) than for the proposed I-DCCRN-VAE (systems (4) and (5)). This can be explained by the difference in the estimated clean speech between DCCRN-VAE and I-DCCRN-VAE before CVAE fine-tuning. Due to posterior collapse, the speech quality of the DCCRN-VAE is rather poor before fine-tuning; the discriminator can easily distinguish between estimated speech and true clean speech, making adversarial training highly effective. In contrast, since the proposed I-DCCRN-VAE learns informative pretrained latent spaces and already produces high-quality speech before fine-tuning, adversarial training does not bring a large benefit. This is a significant advantage, as classical fine-tuning avoids the convergence and sensitivity issues that are common to adversarial training.

Finally, we compare the speech enhancement performance of the proposed I-DCCRN-VAE (systems (4) and (5)) with DCCRN (system (1)) and DCCRN-VAE using adversarial training (system (3)). On the matched DNS3 dataset, it can be observed that the baseline DCCRN and DCCRN-VAE achieve slightly better SI-SDR or PESQ scores than the I-DCCRN-VAE. However, for both considered mismatched datasets, the proposed I-DCCRN-VAE consistently achieves the best performance in terms of all metrics. Specifically, on the WSJ0-QUT dataset, the I-DCCRN-VAE improves upon baselines by around 1.7\,dB in SI-SDR, 0.1 in PESQ, and 0.03 in ESTOI. On the VB-DMD dataset, the respective improvements are 0.6\,dB in SI-SDR, around 0.06 in PESQ, and 0.02 in ESTOI. This demonstrates the improved generalization ability of the proposed I-DCCRN-VAE.

\section{Conclusions}
\label{sec:conclusion}
This paper proposed the I-DCCRN-VAE for complex VAE-based single-channel speech enhancement, which improves the DCCRN-VAE. We demonstrate that three key modifications are crucial for improvement: 1) removing skip connections in the pretrained VAEs to avoid posterior collapse; 2) using $\beta$-VAE in pretrained VAEs to better balance reconstruction and latent space regularization; and 3) the NSVAE generating both speech and noise representations for better speech extraction. Experiments show that the I-DCCRN-VAE achieves comparable performance to baselines on the matched dataset but consistently better performance on two mismatched datasets, demonstrating better generalization. Especially, the I-DCCRN-VAE achieves the performance even with classical fine-tuning, not adversarial training, leading to  a simplified training pipeline.



\bibliographystyle{IEEEbib}
\bibliography{refs}

\end{document}